\newif\iffigs\figstrue
\documentclass[12pt]{article}
\usepackage{latexsym,amssymb}
\iffigs
             \input{epsf}
\else
\fi
\def\IC{\relax\,\hbox{$\inbar\kern-.3em{\rm C}$}}
\def\IG{\relax\,\hbox{$\inbar\kern-.3em{\rm G}$}}
\def\IB{\relax{\rm I\kern-.18em B}}
\def\ID{\relax{\rm I\kern-.18em D}}
\def\IL{\relax{\rm I\kern-.18em L}}
\def\IF{\relax{\rm I\kern-.18em F}}
\def\IH{\relax{\rm I\kern-.18em H}}
\def\II{\relax{\rm I\kern-.17em I}}
\def\IN{\relax{\rm I\kern-.18em N}}
\def\IP{\relax{\rm I\kern-.18em P}}
\def\IQ{\relax\,\hbox{$\inbar\kern-.3em{\rm Q}$}}
\def\bfzero{\relax\,\hbox{$\inbar\kern-.3em{\rm 0}$}}
\def\IK{\relax{\rm I\kern-.18em K}}
\def\IG{\relax\,\hbox{$\inbar\kern-.3em{\rm G}$}}
 \font\cmss=cmss10 \font\cmsss=cmss10 at 7pt
\def\IR{\relax{\rm I\kern-.18em R}}
\def\ZZ{\relax\ifmmode\mathchoice
{\hbox{\cmss Z\kern-.4em Z}}{\hbox{\cmss Z\kern-.4em Z}}
{\lower.9pt\hbox{\cmsss Z\kern-.4em Z}} {\lower1.2pt\hbox{\cmsss
Z\kern-.4em Z}}\else{\cmss Z\kern-.4em Z}\fi}
\def\bfone{\relax{\rm 1\kern-.35em 1}}

\def\inbar{\vrule height1.5ex width.4pt depth0pt}
\def\bfzero{\relax{\rm I\kern-.18em 0}}
\def\bfone{\relax{\rm 1\kern-.35em 1}}

\newcommand{\ft}[2]{{\textstyle\frac{#1}{#2}}}
\def\1bar{1\hskip -.275cm -}
\def\2bar{2\hskip -.275cm -}
\def\3bar{3\hskip -.275cm -}

\newsavebox{\uuunit}
\sbox{\uuunit}
                 {\setlength{\unitlength}{0.825em}
                      \begin{picture}(0.6,0.7)
                                      \thinlines
                                      \put(0,0){\line(1,0){0.5}}
                                      \put(0.15,0){\line(0,1){0.7}}
                                      \put(0.35,0){\line(0,1){0.8}}
                                     \multiput(0.3,0.8)(-0.04,-0.02){10}{\rule{0.5pt}{0.5pt}}
                      \end {picture}}

\makeatletter \@addtoreset{equation}{section} \makeatother
\newcommand{\be}{\begin{equation}}
\newcommand{\ee}{\end{equation}}
\newcommand{\ba}{\begin{eqnarray}}
\newcommand{\ea}{\end{eqnarray}}
\def\bfone{\relax{\rm 1\kern-.35em 1}}

\def\bfone{\relax{\rm 1\kern-.35em 1}}
\font\cmss=cmss10 \font\cmsss=cmss10 at 7pt
\newcommand{\p}[1]{(\ref{#1})}
\newcommand{\nn}{\nonumber}
\begin{document}
\begin{titlepage}
\begin{flushright}
DFTT05/31\\
{hep-th/0510156}
\end{flushright}
\vskip 0.5cm
\begin{center}
{\LARGE {\bf Integrability of Supergravity Billiards
}}\\[0.3cm]
{\LARGE {\bf and
}}\\[0.3cm]
{\LARGE {\bf
the generalized Toda lattice  equations$^\dagger$}}\\[1.cm]
%{\LARGE {\bf    }}\\[1cm]
{\large P. Fr\'e$^{a}$ and A.S. Sorin$^{b}$}
{}~\\
\quad \\
{{\em $^{a}$ Dipartimento di Fisica Teorica, Universit\'a di Torino,}}
\\
{{\em $\&$ INFN - Sezione di Torino}}\\
{\em via P. Giuria 1, I-10125 Torino, Italy}~\quad\\
{\tt fre@to.infn.it}
{}~
{}~
{}~\\
\quad \\
{{\em $^{b}$ Bogoliubov Laboratory of Theoretical Physics,}}\\
{{\em Joint Institute for Nuclear Research,}}\\
{\em 141980 Dubna, Moscow Region, Russia}~\quad\\
{\tt sorin@theor.jinr.ru}
{}~
{}~
\quad \\
\end{center}
~{}
\begin{abstract}
We prove that the field equations of supergravity for purely time-dependent
backgrounds, which reduce to those of a one--dimensional sigma model, admit a Lax pair
representation and are fully integrable. In the case where the
effective sigma model is on a maximally split non--compact coset
$\mathrm{U/H}$ (maximal supergravity or subsectors of lower
supersymmetry supergravities) we are also able to construct a
completely explicit analytic integration algorithm, adapting a method
introduced by Kodama et al in a recent paper. The properties of the general integral are
particularly suggestive. Initial data are represented by a pair
$\mathcal{C}_0,h_0$ where $\mathcal{C}_0$ is in the CSA of the Lie algebra of
$\mathrm{U}$ and $h_0 \in \mathrm{H}/\mathcal{W}$, is in the compact subgroup
$\mathrm{H}$ modded by the Weyl group of $\mathrm{U}$. At asymptotically early and
asymptotically late times the Lax operator is always in the Cartan
subalgebra and due to the iso-spectral property the two limits differ
only by the action of some element of the Weyl group. Hence the
entire cosmic evolution can be seen as a  billiard scattering with
quantized angles defined by the Weyl group. The solution algorithm
realizes a map from $\mathrm{H}/\mathcal{W}$ into $\mathcal{W}$.
\end{abstract}
\vfill
\vspace{1.5cm}

\vspace{2mm} \vfill \hrule width 3.cm {\footnotesize $^ \dagger $
This work is supported in part by the European Union RTN contract
MRTN-CT-2004-005104 and by the Italian Ministry of University (MIUR) under contract PRIN 2003023852}
\end{titlepage}
\section{Introduction}
\paragraph{String Cosmology}
\par
The cosmological implications of superstring theory have been under
attentive consideration in the last few years from various
viewpoints \cite{cosmicstringliteraturegeneral}.
This involves the classification and the study of possible time-evolving string
backgrounds which amounts to the construction, classification and analysis of supergravity solutions
depending only on time or, more generally, on a low number of coordinates including time.
\par
\paragraph{The idea of cosmic billiards}
\par
In this context a quite challenging and potentially
highly relevant phenomenon for the overall interpretation of extra--dimensions and string dynamics is provided by the so
named \textit{cosmic billiard} phenomenon
\cite{cosmicbilliardliterature1},\cite{cosmicbilliardliterature2},\cite{cosmicbilliardliterature3},
\cite{cosmicbilliardliterature4}.
This is based on the relation between the  cosmological scale factors
and the  duality groups $U$ of  string theory. The group
$\mathrm{U}$ appears as isometry group of the scalar
manifold $\mathcal{M}_{scalar}$ emerging in compactifications of
$10$--dimensional supergravity to lower dimensions $D<10$ and
 depends both on the geometry of the compact dimensions and on
the number of preserved supersymmetries $N_Q \le 32$.  For $N_Q > 8$ the scalar manifold is always a homogeneous
space $\mathrm{U/H}$. The cosmological scale factors $a_i(t)$ associated with the
various  dimensions of supergravity  are interpreted as exponentials of those
scalar fields $h_i(t)$ which lie in the Cartan subalgebra of
$\mathbb{U}$, while the other scalar fields in $\mathrm{U/H}$ correspond to
positive roots $\alpha >0$ of the Lie algebra $\mathbb{U}$.
The cosmological evolution is described by  a
\textit{fictitious ball} that moves in the CSA of $\mathbb{U}$ and occasionally bounces on the hyperplanes orthogonal
to the various roots: the billiard walls. Such bounces  represent
inversions in the time evolution of scale factors.
Such a scenario was introduced by Damour, Henneaux, Julia and Nicolai in a
series of papers that we already quoted
\cite{cosmicbilliardliterature1},\cite{cosmicbilliardliterature2},
\cite{cosmicbilliardliterature3},\cite{cosmicbilliardliterature4},
which generalize classical results obtained in the context of pure
General Relativity \cite{Kassner}. In this approach the cosmic billiard
phenomenon is analyzed as an asymptotic regime in the neighborhood of
space-like singularities and the billiard walls are seen as delta
function potentials provided by the various $p$--forms of
supergravity localized at sharp instants of time.
%%%%%%%%%
\paragraph{The smooth cosmic billiard programme}
In a series of papers \cite{noiconsasha},\cite{Weylnashpaper,noipaintgroup,noiKacmodpaper} involving both
the present authors and other collaborators it was started and developed what
can be described as the \textit{smooth cosmic billiard programme}. This
amounts to the study of the \textit{billiard features}, namely
inversions in the evolution of cosmological scale factors within the
framework of exact analytic solutions of supergravity rather than in
asymptotic regimes. Crucial starting point in this programme
was the observation \cite{noiconsasha} that the fundamental
mathematical setup underlying the appearance of the billiard
phenomenon is the so named \textit{Solvable Lie algebra
parametrization} of supergravity scalar manifolds, pioneered in \cite{primisolvi}
and later applied to the solution of a large variety of superstring/supergravity
problems \cite{noie7blackholes},\cite{otherBHpape},\cite{gaugedsugrapot} (for
a comprehensive review see \cite{myparis}).
Thanks to the solvable parametrization, one can
establish a precise algorithm to implement the following programme:
\begin{enumerate}
  \item Reduce the original supergravity in higher  dimensions $D \ge 4$ (for instance $D=10,11$) to
  a gravity-coupled $\sigma$--model in $D \le 3$ where gravity is
  non--dynamical and can be eliminated. The target manifold is the non compact coset
  $\mathrm{U/H} \cong \exp \left[ Solv\left(\mathrm{U/H} \right)
  \right]$ metrically equivalent to a solvable group manifold.
  \item Utilize the  structure  of the algebra
  $Solv\left(\mathrm{U/H} \right) $ in order to integrate analytically the
  $\sigma$--model equations.
  \item Dimensionally oxide the solutions obtained in this way to
  exact  time dependent solutions of  $D \ge 4$ supergravity.
\end{enumerate}
Within this approach it was proved in \cite{noiconsasha} that the
\textit{cosmic billiard phenomenon} is indeed a general feature of exact time
dependent solutions of supergravity and has \textit{smooth
realizations}. Calling $\mathbf{h}(t)$ the $r$--component
vector of Cartan fields (where $r$ is the rank of $\mathbb{U}$) and
$\mathbf{h}_\alpha(t) \equiv \mathbf{\alpha} \, \cdot \,
 {\mathbf{h}}(t)$ its projection along any positive root $\alpha$,
a \textit{bounce} occurs at those instant of times $t_i$ such that:
\begin{equation}
\exists \,  \mathbf{\alpha} \, \in \, \Delta_+ \, \quad  \backslash
\quad \quad
 \dot{\mathbf{h}}_\alpha (t)\mid_{t=t_i} \, = \, 0
\label{bouncino}
\end{equation}
namely when the Cartan field in the direction of some root $\alpha$ inverts
its behaviour. All higher dimensional
bosonic fields (off-diagonal components of the metric $g_{\mu\nu}$ or
$p$--forms $A^{[p]}$) are, via the solvable parametrization of $\mathrm{U/H}$, in one-to-one correspondence with roots
$ \phi_\alpha \, \Leftrightarrow \, \alpha$, and each bounce on a hyperplane orthogonal to a root $\alpha$ is
caused by the sudden growing of the field $\phi_\alpha$ which, in exact smooth
solutions, has a typical bell-shaped behaviour  around the bouncing time $t=t_b$.
\par
\subsection{Algebraic structure of $\mathrm{U/H}$ and integration with the compensator method}
As alluded to in the above discussion, the actual construction of
exact analytic solutions performed in \cite{noiconsasha} and
\cite{Weylnashpaper} was based on two ingredients, namely the
systematics of \textit{dimensional reduction/oxidation} which ultimately allows
to reduce supergravity field equations to those of a one--dimensional sigma
model on a coset $\mathrm{U/H}$ and secondly the algebraic structure
of $\mathrm{U/H}$ which allows to device effective integration
methods. The essential point relative to the second ingredient is
that in all supergravity models $\mathrm{U/H}$ is non--compact and
for maximal supersymmetry it is also \textit{maximally split}, namely the Lie
algebra $\mathbb{U}$ of $\mathrm{U}$ is the maximally non compact real section of its own
complexification and $\mathrm{H}$ is generated by the maximal compact
subalgebra $\mathbb{H}\subset \mathbb{U}$. In this case the
generators $t_\alpha$ of $\mathbb{H}$ are in one-to-one correspondence with
the positive roots $\alpha$ of $\mathbb{U}$ and the detailed algebraic
structure of this geometry was utilized in \cite{noiconsasha} to
device an integration algorithm named  \textit{the compensator
method}. This latter essentially consists of substituting the
original $\sigma$--model field equations with an equivalent set of
differential equations which are more manageable since they can be put
into triangular form and integrated one--by--one. The geometric meaning of
these new equations is that they correspond to the conditions to be
satisfied by the $\theta_\alpha(t)$ parameters of an $\mathrm{H}$--gauge transformation
$ \mathbb{L} \,\mapsto \, \mathbb{L} h(\theta) $
in order for  the so called solvable gauge of the
$\mathrm{U/H}$ coset representative $\mathbb{L}(t)$ to be preserved,
solvable gauge being the possibility of writing $\mathbb{L}(t)$ as
an exponential of the solvable Lie algebra $Solv(\mathrm{U/H})$.
The limitations of the approach developed in \cite{noiconsasha,Weylnashpaper} and of
the compensator method which nonetheless was very valuable in producing a rich set of explicit analytic solutions
are threefold:
\begin{description}
  \item[a] In its original form, the scheme of \cite{noiconsasha,Weylnashpaper} applies
  only to \textit{finite  algebras}, namely to the dimensional
  reduction of supergravity to $D=3$, which is the first where all
  (Bose) degrees of freedom can be represented by scalar fields.
  This is not yet sufficient to reveal the full algebraic structure
  underlying time evolution and billiard dynamics. Indeed to discuss this latter one can
  still dimensionally reduce to $D=2$ and $D=1$ where the algebra
  $\mathbb{U}$ is extended to an affine or hyperbolic Ka\v c--Moody
  algebra. It is therefore mandatory to consider the field
  theoretical mechanisms of such Ka\v c--Moody extensions and
  subsequently extend the so far considered algebraically based integration
  methods to Ka\v c--Moody algebras.
  \item[b] Heavily relying on the \textit{maximally split} character
  of the pair $\left \{  \mathbb{U} \, , \, \mathbb{H}\right\} $ the
  original form of the scheme presented in
  \cite{noiconsasha,Weylnashpaper} is not directly applicable
  to the case of lower supersymmetry $N_Q < 32$ and hence to
  more realistic and more interesting compactifications than pure toroidal
  ones.
  \item[c] The compensator method, although very valuable in
  obtaining explicit solutions, replaces one differential system with
  another one and therefore does not answer the question whether the
  system is a fully integrable system nor it is able to provide  a
  uniformly given, general integral, depending on as many integration
  constants as necessary. Indeed the solution strategy is based on
  triangulization  of the set of equations and the choice of parametrization of
  the $\mathrm{H}$-subgroup element $h$ (order of the rotations
  $\exp\left [ \theta_\alpha(t) t_\alpha\right ]$ )
  in such a way that the resulting differential system for the rotation angles $\theta_\alpha(t)$
  be triangular is something that has to be decided case by case and both by intuition and by trial and error.
\end{description}
The two recent papers \cite{noiKacmodpaper} and \cite{noipaintgroup}
have presented some substantial advances with respect to points $\mathbf{a}]$
and $\mathbf{b]}$ of the above list.
\paragraph{$\mathbf{a}]$: In \cite{noiKacmodpaper}} it was
clarified the general field theoretical mechanism underlying Ka\v
c--Moody extensions in supergravity and the general pattern of such
extensions in the reduction to $D=2$ of all $D=4$ supergravity theories with
all number of supercharges $N_Q$ was also presented. This means that
such extensions are now under control for all compactifications of both
M-theory and string theory on all type of compact geometries. Indeed these latter
lead to the various different types of $D=4$ supergravities. Although integration methods for
the Ka\v c-Moody analogue of  the $\sigma$--model equations
were not presented in \cite{noiKacmodpaper}, yet the results there obtained
provide the necessary prerequisite in the development of such a
programme. Indeed the field theoretical mechanism of the extension is
what one needs to consider in order to rewrite the actual $D=2$ or
$D=1$ field equations in a manifestly Ka\v c-Moody covariant way.
\par
\paragraph{$\mathbf{b}]$: In \cite{noipaintgroup},} instead, exploiting in a systematic way the Tits
Satake theory of real forms \cite{Helgason} for simple algebras and the Tits-Satake projection of
root--spaces, a decisive advance was made with respect to point
$\mathbf{b]}$ of the above list. Indeed it was shown how the
differential equations of the $\sigma$--model on \textit{a non maximally
split non compact coset} $\mathrm{U/H}$, typically emerging from non
maximal supergravities, can be consistently reduced to those of a
$\sigma$-model on a \textit{maximally split one}: $\mathrm{U_{TS}/H_{TS}}$.
By $\mathbb{U}_{TS} \subset \mathbb{U}$ we denote the Tits-Satake
subalgebra of the original algebra $\mathbb{U}$ which is uniquely defined by
the involutive automorphism $\sigma$ associated with the real form of
$\mathbb{U}$.
By $\mathbb{H}_{TS}\subset \mathbb{U}_{TS}$ we denote
the maximal compact subalgebra of the Tits Satake subalgebra which, as in all maximally split cases, has a number of
generators equal to the number of positive roots in the   $\overline{\Delta}_{\mathbb{U}}$ root system of
$\mathbb{U}_{TS}$. This latter is the Tits Satake  projection of the $\mathrm{U}$--root system:
\begin{equation}
  \Pi \quad : \quad \Delta_\mathbb{U} \, \mapsto  \,
  \overline{\Delta}_\mathbb{U}
\label{projecto}
\end{equation}
obtained by setting to zero all transverse components $\alpha_\bot$
of each root $\alpha$. Indeed the automorphism $\sigma^2=\mathbf{1}$
defining the real form of $\mathbb{U}$ splits the Cartan subalgebra
$\mathrm{CSA} \subset \mathbb{U}$ in two eigenspaces $\mathcal{H}^{n.c.}$ and
$\mathcal{H}^{comp}$,
respectively corresponding to the eigenvalues $\pm 1$. Each root
$\alpha$, which is a linear functional  on the $\mathrm{CSA}$, is
accordingly decomposed into a parallel and into a transverse part to the
non compact Cartan subalgebra $\mathcal{H}^{n.c.}$:
\begin{equation}
  \alpha = \alpha_{||} \, \oplus \, \alpha_{\bot} \quad ; \quad
  \alpha_{\bot}\left( \mathcal{H}^{n.c.}\right) = 0 \quad ; \quad
  \alpha_{||}\left( \mathcal{H}^{n.c.}\right) \ne 0.
\label{splittus}
\end{equation}
The Tits Satake projection projects each root $\alpha$ onto its
parallel part $\alpha_{||}$.
\par
As already mentioned it was shown in \cite{noipaintgroup} that in the \textit{non maximally
split cases} one can consistently reduce the $\sigma$--model equations of
$\mathrm{U/H}$ to those of $\mathrm{U_{TS}/H_{TS}}$ and furthermore
it was proved that one can lift back any solution of these latter to
a full fledged solution of the original system by means of the action
of an automorphism compact group named the paint group $\mathrm{G}_{paint}
\subset \mathrm{U}$ with respect to which the entire solvable algebra
$Solv_{\mathbb{U}}$ transforms as a representation:
\begin{eqnarray}
\left[ \mathbb{G}_{paint} \, , \, Solv_{\mathbb{U}} \right]  =
   Solv_{\mathbb{U}}.
\label{linearrepre}
\end{eqnarray}
Hence a large class of solutions of the $\sigma$--model on
\textit{non--maximally split coset manifolds}, and presumably the
essentially relevant class for billiard dynamics, since it contains
all the Cartan fields and all the walls on which the cosmic ball can
bounce, can be obtained from the general solution of $\sigma$--models
on \textit{maximally split cosets}.
\par
\subsection{The present paper solves point $\mathbf{c}]$: full integrability
via LAX representation}
\label{anticipazia} In the present paper
we shall  present a complete solution for the third point in the above list of problems
and required generalizations. Indeed we shall prove that the
one--dimensional $\sigma$--model on \textit{maximally split cosets}
is completely integrable in the sense of integrable systems,  since
the first order equations for the tangent vector to a geodesic can be
recast in the classical form of a LAX system. Furthermore, adapting
to our case an algorithm developed  in \cite{kodama}, we can
write a closed analytic form for the general integral, depending on a
complete set of integration constants.
\par
This new algorithm replaces the compensator method and does not
require the solution of neither differential nor algebraic equations.
The solution of the equations for the time parameter ranging from
$-\infty$ to $+\infty$ is directly constructed in terms of the
initial data fixed say at $t=0$. It is particularly interesting to
study the general properties of the solution algorithm. As we shall explain, the initial
data parametrizing the value $L_0$ of the Lax operator $L(t)$ at $t=0$ can be represented as a pair:
\begin{enumerate}
  \item An element of the Cartan subalgebra
  $\mathcal{C}_0 \in \mathrm{CSA} \subset \mathbb{U}$.
  \item An element $h_0 \in \mathrm{H} = \exp\left [\mathbb{H} \right]$ of
  the maximally compact subgroup $\mathrm{H}\subset \mathrm{U}$.
\end{enumerate}
The solution algorithm $\mathcal{S}_K$ provides a general integral
\begin{equation}
  \mathcal{S}_K \, : \, \left\{\mathcal{C}_0 \, , \,h_0 \right\}
  \, \Longrightarrow \, L\left ( t \, | \, \mathcal{C}_0  ,  h_0\right)
\label{geninte}
\end{equation}
with the following properties:
\begin{enumerate}
  \item {$ \forall \; w \, \in \, \mathcal{W} $ where $ \mathcal{W} \,
  \subset \, \mathrm{H} \, \subset \, \mathrm{U}$ is the Weyl group
  of the Lie algebra $\mathbb{U}$
\begin{equation}
  L\left ( t \, | \, \mathcal{C}_0  , w \cdot h_0\right) \, = \, L\left
  ( t \, | \, w^{-1}\cdot\mathcal{C}_0 \cdot
   w ,  h_0\right).
\label{Lweylproperty}
\end{equation}
}
\item {The limits of the Lax operator $ L\left ( t \, | \, \mathcal{C}_0  ,
  h_0\right)$ at $t=\pm \infty$ lie in the Cartan subalgebra:
\begin{equation}
  \lim_{t\rightarrow \pm \infty} \, L\left ( t \, | \, \mathcal{C}_0  , h_0\right) \, = \,
  L_{\pm\infty} \left( \mathcal{C}_0  , h_0\right) \, \in \, \mathrm{CSA}.
\label{pminftylimits}
\end{equation}}
\item{At any instant of time the eigenvalues $\left \{\lambda_1 \, \lambda_2
\, \dots \, \lambda_n \right \}$ of the Lax operator are the same and
are given by
\begin{equation}
  \lambda_i = \mathbf{w}_i \left( \mathcal{C}_0 \right)
\label{pesini}
\end{equation}
where $\mathbf{w}_i$ are the weights of the representation of
$\mathbb{U}$ to which the Lax operator is assigned.
}
\end{enumerate}
Property (2) and property (3) combined together imply that the two
asymptotic values $L_{\pm\infty}\left ( \mathcal{C}_0  , h_0\right)$ of
the Lax operator are necessarily related to each other by some
element  $\Omega \in \mathcal{W}$ of the Weyl group which represents
a sort of topological charge of the solution:
\begin{equation}
\exists \; \Omega \, \in \, \mathcal{W} \, :  \,   L_{+\infty}\left (  \mathcal{C}_0  , h_0\right) =
\Omega^{-1} \, \cdot \, L_{-\infty}\left (  \mathcal{C}_0  , h_0\right) \, \cdot \, \Omega~.
\label{topcharge}
\end{equation}
Indeed since the Weyl group is discrete, by varying the initial
datum $h_0$  a solution which is characterized by a
\textit{charge} $\Omega_1$, cannot continuously be deformed into
another characterized by a different \textit{charge} $\Omega_2$.
\par
It happens furthermore that the topological charge $\Omega$ of a
solution is independent of  the choice of $\mathcal{C}_0 \in
\mathrm{CSA}$ and just depends on $h_0$. We conclude that the
solution algorithm $\mathcal{S}_K$ developed in \cite{kodama} induces a map:
\begin{equation}
  \mathcal{P}_K \, : \, \frac{\mathrm{H}}{\mathcal{W}} \, \rightarrow
  \, \mathcal{W} \, \equiv \, \pi_1 \left( \frac{\mathrm{H}}{\mathcal{W}} \right )
\label{Pmappa}
\end{equation}
from the non--simply connected coset manifold
$\frac{\mathrm{H}}{\mathcal{W}}$ to its own homotopy group.
\par
Let us now proceed to the derivation of these results and to the
illustration of their application to cosmic billiards.
\section{One-dimensional sigma models on
maximally split coset manifolds and Lax pairs}
In view of the discussion presented in the introduction our
mathematical object of study is reduced to be the one-dimensional
$\sigma$-model on a maximally split hermitian symmetric coset
manifold $\mathrm{U/H}$. The corresponding action principle is given
by:
\begin{eqnarray} \label{act}
S=\int dt \ h_{IJ}(\phi) \frac{d\phi}{dt}^I\frac{d\phi}{dt}^J
\end{eqnarray}
where $h_{IJ}$ is the canonical metric on the coset, which we parameterize by scalar fields named $\phi^I$.
By definition, the coset $\mathrm{U}/\mathrm{H}$ is hermitian symmetric if the Lie algebra $\mathbb{U}$ of
the group $\mathrm{U}$ and the subalgebra $\mathbb{H}$ of the subgroup
$\mathrm{H}$ fulfill the relations:
\begin{eqnarray} \label{sym}
 [\mathbb{H},\mathbb{H}]
\subset\mathbb{H},\ \ \ [\mathbb{H},\mathbb{K}]
\subset\mathbb{K},\ \ \ [\mathbb{K},\mathbb{K}]
\subset\mathbb{H}\ \ \
\end{eqnarray}
having called $\mathbb{K}$ the orthogonal complement of $\mathbb{H}$ in
$\mathbb{U}$, namely:
\begin{eqnarray}
\mathbb{H}\subset \mathbb{U}, \quad
\mathbb{U}= \mathbb{H}\oplus \mathbb{K}~.
\label{cosetdecompo}
\end{eqnarray}
In order to asset the further property of being maximally split we
assume that $\mathbb{U}$ is a \textit{simple Lie algebra} in its maximally non-compact
real section and we normalize the Cartan--Weyl commutation relation in the standard form:
\begin{eqnarray}
\left[ \mathcal{H}_i \, , \, \mathcal{H}_j\right ]  & =  & 0,\nonumber\\
\left[ \mathcal{H}_i \, , \, E^\alpha \right]  & =  & \alpha^i \, E^\alpha, \nonumber\\
\left[ E^\alpha \, , \, E^{-\alpha} \right]  & =  & \alpha \, \cdot \,  \mathcal{H},
\nonumber\\
\left[ E^{\alpha_i} \, , \, E^{\alpha_j} \right]  & =  &
\mathcal{N}_{\alpha_i , \alpha_j} \, E^{\alpha_i + \alpha_j}
\label{cartaweila}
\end{eqnarray}
where $\mathcal{H}_i$ denotes a basis for the Cartan subalgebra
(CSA), $\alpha_i$ the root vectors of the root system $\Delta$ and
$E^{\alpha_i}$ the corresponding step operators. Then $\mathrm{U}/\mathrm{H}$ is
maximally split if the subalgebra $\mathbb{H}$ is chosen to be the maximally
compact subalgebra which has  dimension equal to the number of
positive roots and which can be defined as:
\begin{eqnarray}
\mathbb{H} = \mathrm{Span}\{t_{\alpha}\} =
\mathrm{Span}\{(E^{\alpha} - E^{-\alpha})\}~.
\label{Hmaxi}
\end{eqnarray}
Eq. (\ref{Hmaxi}) defines the subalgebra and also a basis for it given
by the generators $t_{\alpha} \equiv (E^{\alpha} - E^{-\alpha})$. In
this case the complementary space $\mathbb{K}$ is given by
\begin{eqnarray}
&& \mathbb{K}=\mathrm{Span}\{\mathbb{K}_A\} = \mathrm{Span}\{\mathcal{H}_i,
\frac{1}{\sqrt{2}}(E^{\alpha} + E^{-\alpha})\}
\end{eqnarray}
which, similarly to eq.(\ref{Hmaxi}) introduces a canonical basis of generators  $\mathbb{K}_A$.
\par
For any parametrization of the coset provided by a coset representative $\mathbb{L}(\phi)$ one can write
down the $\mathbb{U}$--algebra valued left invariant one-form
\begin{eqnarray} \label{Omega} \Omega=\mathbb{L}^{-1}\frac{d}{dt} \mathbb{L} =
W^\alpha t_\alpha +V^A \mathbb{K}_A \equiv W+V
\end{eqnarray}
where $V$ is the coset manifold vielbein using which one can
rewrite action (\ref{act}) as follows
\begin{eqnarray}
S=\int d t\ \mathrm{Tr} \left(V V \right).
\label{tracerepresentation}
\end{eqnarray}
In eq. (\ref{tracerepresentation}) it is understood that we have
chosen a linear representation of the Lie algebra $\mathrm{U}$ (which one is
irrelevant, typically the lowest dimensional one) and therefore the
vielbein has become a $ n \times n $ matrix valued one--form, $n$
being the dimension of the chosen representation:
\begin{equation}
  V (t) = \mathbb{K}_A \, V^{A}_{\phantom{A}I}\left(\phi(t) \right)
  \, \frac{d\phi^I}{dt} = \mathbb{K}_A \, V^A~.
\label{explicitV}
\end{equation}
Our goal is to solve the classical field equation of the model
(\ref{act}), namely calculate the geodesics of the manifold
$\mathrm{U/H}$.
Let us calculate the variation of the action
\begin{eqnarray} \label{varS}
\delta S=2 \int d t\ \mathrm{Tr} \left(V \delta V \right).
\end{eqnarray}
In order to derive a convenient expression for the variation of the vielbein
$\delta V$ which enters eq. \p{varS}, we represent the variation of the
left--invariant one-form $\Omega$
defined in eq. \p{Omega} in the following equivalent way:
\begin{eqnarray}
\label{equiv}\delta \Omega=[\Omega, \mathbb{L}^{-1}\delta
\mathbb{L}]+\frac{d}{dt}(\mathbb{L}^{-1}\delta \mathbb{L} )
\end{eqnarray}
and take into account that
\begin{eqnarray} \label{rel} \mathbb{L}^{-1}\delta \mathbb{L}=\delta \omega^\alpha
t_\alpha+\delta \upsilon^A \mathbb{K}_A\equiv\delta\omega+\delta\upsilon,
\end{eqnarray}
where $\delta\upsilon^I$ and $\delta \omega^\alpha$ are functionals of
the variation of the fields $\delta \phi^I$, and there is the one-to-one
correspondence between $\delta\upsilon^I$ and $\delta \phi^I$ (and between
$\upsilon^I$ and $\phi^I$ modulo trivial constants), so that the transition
from $\delta \phi^I$ to $\delta\upsilon^I$ can be treated as a change of the basis.
 Substituting the relations \p{Omega} and \p{rel} into eq. \p{equiv}
we obtain the following new relation:
\begin{eqnarray} \label{varOm}\delta W+\delta V=[W+
V,\delta\omega+\delta\upsilon ]+
\frac{d}{dt} (\delta\omega+\delta\upsilon ).
\end{eqnarray}
Now, using eq. \p{sym} it is easy to extract from
eq. \p{varOm} the projection along the coset generators $\mathbb{K}_A$ and derive
$\delta V$
 \begin{eqnarray} \delta V=[W,\delta\upsilon ]+[V,\delta\omega ]+
\frac{d}{dt} \delta\upsilon.
\end{eqnarray}
Substituting this expression into \p{varS} one can easily obtain
the variation of the action $\delta S$
\begin{eqnarray}
\delta S&=&2\int dt\  \mathrm{Tr} \left( V[W,\delta\upsilon]
+V[V,\delta\omega ]+V \frac{d}{dt} \delta\upsilon \right)\nonumber\\
&=&2\int dt\ \mathrm{Tr}\left ( \left([V,W ]-\frac{d}{dt} V \right)
\delta\upsilon\right).
\end{eqnarray}
The equations of motion follow immediately from the vanishing of the
action variation $\delta S = 0$
\begin{eqnarray} \label{LaxEq}
 \frac{d}{dt} V=\left [V\, , \, W \right]
 \label{Laxequazia}
\end{eqnarray}
in the form of a Lax pair representation. Hence the resulting system
is indeed integrable. The direct consequence of the Lax pair representation
\p{LaxEq} is the existence of the following integrals of motion:
 \begin{eqnarray} \label{integrals}
  \mathcal{I}_k=\mathrm{Tr} \, \left( V^k\right).
 \end{eqnarray}
for all integers $k \le n$, having called $n$ the dimension of the
matrix $V$ namely the dimension of the chosen linear representation
of the Lie algebra $\mathbb{U}$, as we have already specified.
\subsection{Nomizu connection for maximally split cosets and the Lax pair
representation}
\label{firstorder}
Now in order to make contact with our previous work and with the
discussion of cosmic billiards, we want to compare the Lax pair
representation of the field equations (\ref{Laxequazia}) with the
form of the same equations derived in \cite{noiconsasha} by means of
the Nomizu connection on solvable Lie algebras.
There  showed that the field equations of the purely
time dependent $\sigma$-model (\ref{act}), which is what we are supposed to solve
in our quest for  time dependent solutions of supergravity, can be
written as follows:
\begin{equation}
\label{D=3feqn} \dot{Y}^A \,+\, \Gamma^A_{BC} Y^B Y^C
\,=\,0
\end{equation}
where ${Y}^A$ denotes the purely time dependent tangent vectors to the geodesic  in an anholonomic
basis:
\begin{equation}
Y^A \,=\,  \cases{Y^i = V^i_I\left(\phi\right) \dot{\phi}^I  \quad \quad \quad i\in \mbox{CSA} \cr
Y^\alpha = \sqrt{2} \, V^\alpha_I\left(\phi\right) \dot{\phi}^I   \quad \alpha\in \mbox{positive root system
$\Delta_+$}\cr}
\label{anholotang}
\end{equation}
$V^A_I\left(\phi\right) d{\phi}^I$ being the vielbein  of
the target manifold we are considering, namely the same object already introduced in the previous section.
In eq. (\ref{D=3feqn}) the symbol $\Gamma^A_{BC}$ denotes the components of the Levi--Civita
connection in the chosen anholonomic basis. Explicitly they are
related to the components of the Levi--Civita connection in an
arbitrary holonomic basis by:
\begin{equation}
\Gamma^A_{BC}\,= \Gamma^I_{JK}V^A_I V^J_B V^K_C - \partial_K
(V^A_J)V^J_B V^K_C \label{capindgamma}
\end{equation}
where the inverse vielbein is defined in the usual way:
\begin{equation}
V^A_I \, V^I_B \,=\, \delta^A_B~. \label{invervielb}
\end{equation}
The basic idea of \cite{noiconsasha}, which was exploited together with the \textit{compensator method}
in order to construct
explicit solutions,  is the following. The connection
$\Gamma^A_{BC}$ can be identified with the \textit{Nomizu
connection} defined on a solvable Lie algebra, if the coset
representative $\mathbb{L}$ from which we construct the vielbein
is solvable, namely if it is represented as the exponential of the associated solvable Lie algebra $Solv(\mathrm{U/H})$.
 In fact, as we can read in \cite{alekseevskii} once we
have defined over $Solv$ a non degenerate, positive definite
symmetric form:
\begin{eqnarray}
\langle \,,\, \rangle & \; : \; Solv \otimes Solv \longrightarrow \mathbb{R}, \nonumber \\
\langle X \,,\, Y \rangle & \; = \; \langle Y \,,\, X \rangle
\end{eqnarray}
whose lifting to the manifold produces the metric, the covariant
derivative is defined through the \textbf{Nomizu operator}:
\begin{equation}
\forall X \in Solv \,:\, \mathbb{L}_X: Solv \longrightarrow Solv
\end{equation}
so that
\begin{equation}
\forall X,Y,Z \in Solv \,:\, 2 \langle Z \,,\, \mathbb{L}_X Y
\rangle \,=\, \langle Z, \left[ X,Y \right] \rangle \,-\, \langle
X, \left[ Y,Z \right] \rangle \,-\, \langle Y, \left[ X,Z \right]
\rangle \label{Nomizuoper}
\end{equation}
while the Riemann curvature 2-form is given by the commutator of
two Nomizu operators:
\begin{equation}
R^W_{\phantom{W}Z} \left( X,Y \right) \,=\, \langle W \,,\,
\left\{ \left[ \mathbb{L}_X , \mathbb{L}_Y \right] \,-\,
\mathbb{L}_{\left[X,Y\right]} \right\} Z \rangle ~.
\label{Nomizucurv}
\end{equation}
This implies that the covariant derivative explicitly reads:
\begin{equation}
\mathbb{L}_X \,Y \,=\, \Gamma_{XY}^Z \,Z \label{Gammonedefi}
\end{equation}
where
\begin{equation}
\Gamma_{XY}^Z \,=\,\frac{1}{2}\left( \langle Z, \left[ X,Y \right]
\rangle \,-\, \langle X, \left[ Y,Z \right] \rangle \,-\, \langle
Y, \left[ X,Z \right] \rangle\right) \, \frac{1}{<Z,Z>} \,\qquad\,
\forall X,Y,Z \in Solv ~.\label{Nomizuconne}
\end{equation}
\par
Eq.(\ref{Nomizuconne}) is true for any solvable Lie algebra, but in
the case of \textbf{maximally non-compact, split algebras} we can
write a general form for $\Gamma_{XY}^Z$, namely:
\begin{eqnarray}
\Gamma^i_{jk} & \,=\, & 0, \nonumber \\
\Gamma^i_{\alpha\beta} & \,= \, & \ft 12 \left(-\langle
E_\alpha,\,\left[E_\beta,\,H^i\right]\rangle - \langle
E_\beta,\,\left[E_\alpha,\,H^i\right]\rangle\right)\,=\,  \, \ft 12
\,
\alpha^i \delta_{\alpha\beta},
\nonumber \\
\Gamma^{\alpha}_{ij}  & \,=
\, & \Gamma^{\alpha}_{i\beta} \,=\, \Gamma^i_{j\alpha} \,=\,0, \nonumber \\
\Gamma^{\alpha}_{\beta i} & \,= \, & \frac{1}{2}\left(\langle
E^\alpha,\,\left[E_\beta,\,H_i\right]\rangle -
\langle E_\beta,\,\left[H_i,\,E^\alpha\right]\rangle\right)\,
=\, -\alpha_i\,\delta^{\alpha}_{\beta},\nonumber\\
\Gamma^{\alpha+\beta}_{\alpha\beta}  & \,=
\, &-\Gamma^{\alpha+\beta}_{\beta\alpha}\,=\,\frac{1}{2} N_{\alpha\beta},\nonumber\\
 \Gamma^{\alpha}_{\alpha+\beta\,\beta}  & \,=
 \, & \Gamma^{\alpha}_{\beta\,\alpha+\beta} \,=\,\frac{1}{2} N_{\alpha\beta}
 \label{Nomizuconne2}
\end{eqnarray}
where $N^{\alpha\beta}$ is defined by the commutator
$\left[ E_\alpha \,,\, E_\beta \right] \,=\,
N_{\alpha\beta}\,E_{\alpha + \beta} \label{nalfabeta}$, as usual.
The explicit form (\ref{Nomizuconne2}) follows from the choice of the
non degenerate metric:
\begin{eqnarray}
\langle \mathcal{H}_i \,,\, \mathcal{H}_j \rangle & \,=\, & \, 2 \, \delta_{ij}, \nonumber \\
\langle \mathcal{H}_i \,,\, E_\alpha \rangle & \,=\, & 0, \nonumber \\
\langle E_\alpha \,,\, E_\beta \rangle & \,=\, &
\delta_{\alpha,\beta}
\end{eqnarray}
$\forall \mathcal{H}_i ,\, \mathcal{H}_j \,\in\,
\mathrm{CSA} $ and $\forall E_\alpha$, step
operator associated with a positive root $\alpha \in \Delta_+$.
\par
Hence in the case of maximally split algebras the first order
equations, take the general form:
\begin{eqnarray}
\label{D=3feqn_2} \dot{Y}^i & \,+\, & \ft 12
\,\sum_{\alpha\in \Delta_+}  \alpha^i Y_\alpha^2 \,=\, 0,
\nonumber \\
\dot{Y}^\alpha &\,+\, &  \,\sum_{\beta\in
\Delta_+}\,N_{\alpha\beta} Y^{\beta}  Y^{\alpha+\beta} -
\alpha_i\,Y^i Y^\alpha\,=\,0
\end{eqnarray}
which follows from eq. (\ref{Nomizuconne2}).
\par
Our next point is that equations (\ref{D=3feqn_2}) are identical to
the Lax pair equations (\ref{LaxEq}) if we identify:
\begin{equation}
\begin{array}{rclcl}
V  & = & V^A \, \mathbb{K}_A & \equiv  & Y^i \, \mathcal{H}_i \, + \,
Y^\alpha \, \ft {1}{2} \left( E^\alpha + E^{-\alpha} \right), \\
W & = & W^\alpha \, \mathbb{H}_\alpha & \equiv  & Y^\alpha \, \ft 12 \, \left( E^\alpha - E^{-\alpha}
\right).\
\label{agnuso1}
\end{array}
\end{equation}
On one hand insertion of (\ref{agnuso1}) into (\ref{LaxEq}),  upon use of the standard Cartan Weyl commutation
relations (\ref{cartaweila}), yields (\ref{D=3feqn_2}), on the other
hand eqs. (\ref{agnuso1}) are justified as follows. From
eq.(\ref{anholotang}) we have
\begin{equation}
  V^\alpha = \frac{1}{\sqrt{2}} \, Y^\alpha
\label{VinY}
\end{equation}
while, on the other hand we have that the condition for the coset
representative $\mathbb{ L }(\phi)$ to be solvable is given by:
\begin{equation}
  V^\alpha = \sqrt{2} \,W^\alpha~.
\label{solvocondo}
\end{equation}
Together eqs. (\ref{VinY}) and (\ref{solvocondo}) imply
(\ref{agnuso1}).
In this way we have proved that the first order equations
(\ref{D=3feqn_2}) studied in
\cite{noiconsasha,Weylnashpaper,noipaintgroup} and for which we have
also recently found bouncing solutions \cite{noipaintgroup} also in the interesting case
of $\mathrm{F_{4(4)}/Usp(6)\times SU(2)}$, which is relevant for $N=6$ supergravity,
are actually fully integrable since they admit the Lax pair rewriting
(\ref{Laxequazia}).
In \cite{noiconsasha,Weylnashpaper,noipaintgroup} eqs.
(\ref{D=3feqn_2}) were addressed by means of the compensator method.
Let us recall its form in order to compare it with the general integral
formula provided by the algorithm of \cite{kodama}.
\subsection{Minireview of the compensator method}
As we have seen above,  the condition which ensure that a coset
representative $\mathbb{L}(\phi)$ is solvable, namely that can be
represented as the exponential of the solvable Lie algebra
$Solv(\mathrm{U/H})$ is given by the linear relation
(\ref{solvocondo}) on the components of the left-invariant one--form
(\ref{Omega}). The compensator method streams from the following
argument. Given a solvable coset representative $\mathbb{L}(\phi)$,
one asks the question how many other solvable representatives are
there in the same equivalence class. This amounts to investigating
the condition to be satisfied by the H-gauge transformation
\begin{eqnarray}
  \mathbb{L} & \mapsto & \mathbb{L} \, h = \overline{\mathbb{L}},
  \nonumber \\
   h &=& \exp \, \left[  \theta^\alpha  \, t_\alpha  \right]
\label{hgauga}
\end{eqnarray}
in order for the solvable gauge (\ref{solvocondo}) to be preserved.
This condition is a system of differential equations, namely:
\begin{equation}
 \frac{\sqrt{2}}{\mbox{tr}(t_\alpha ^2)} \, \,\mbox{tr} \left( h^{-1}(\theta)\, dh(\theta) \, t_\alpha
  \right)= V^\beta \, \left( - A(\theta)_\beta^{\phantom{\beta}\alpha}
  \, +\, D(\theta)_\beta^{\phantom{\beta}\alpha} \right) \, + \,  V^i \,
  D(\theta)_i^{\phantom{i}\alpha}.
\label{daequa}
\end{equation}
In eq. (\ref{daequa}) the matrix $A(\theta)$ is the adjoint representation of $h \in \mathrm{H}$
and $D(\theta)$ is the $D$--representation of the same group
element which acts on the complementary space $\mathbb{K}$ and which depends case to
case
\begin{eqnarray}
h^{-1} \, t_\alpha  \, h & = & A(\theta)_\alpha^{\phantom{\alpha }\beta} \,t_\beta , \nonumber\\
h^{-1} \, \mathbb{K}_A  \, h &  = & D(\theta)_A^{\phantom{A }B} \,\mathbb{K}_B~.
\label{duematrici}
\end{eqnarray}
The differential system (\ref{daequa}) is actually equivalent to the
original system (\ref{D=3feqn_2}). To see this it suffices to argue
as follows. A simple solution of the first order equations (\ref{D=3feqn_2}) is
easily obtained by setting $Y^\alpha=0$ and $Y^i=c^i=\mbox{const}$,
namely we can begin with a constant vector in the direction of the $\mathrm{CSA}$.
Such a solution is named the \textit{the normal form} of the tangent
vector. In the language of billiard dynamics it corresponds to a
\textit{fictitious ball} that moves on a straight line with a
constant velocity. All other solutions of eqs. (\ref{D=3feqn_2}) can
be obtained from the normal form solution by means of successive
rotations of the compact group, with parameters $\theta[t]$
satisfying the differential equation (\ref{daequa}).
The advantage of this method, emphasized in \cite{noiconsasha} where we introduced it
is that we can choose rotations
in such a way that at each successive rotation we obtain an equation which is
fully integrable in terms of the integral of the previous ones. How
to do this, however, can not be prescribed in general and furthermore
it is not obvious how many steps one can do of this type. In other
words, although the compensator method provides a valuable tool to
obtain several explicit solutions, it does not suffice to reveal the
complete integrability which is instead revealed by the Lax pair
representation (\ref{Laxequazia}).
\par
\subsection{The $A_2$ case with the compensator method}
To appreciate the bearing of the compensator method relative to the
general integration algorithm that we shall present in next section
let us consider the simplest example of non abelian maximally split
coset, already studied and solved in \cite{noiconsasha}. This is the
manifold
\begin{equation}
  \mathrm{U/H} = \mathrm{SL(3,\mathbb{R})}/\mathrm{SO(3)}
\label{UHA2}
\end{equation}
and the Lie algebra $\mathrm{U=SL(3,\mathbb{R})}$ is the maximally
non-compact real form of the simple complex Lie algebra $A_2$. The
coset decomposition (\ref{cosetdecompo}) takes the form:
\begin{eqnarray}
\mathrm{SL(3,\mathbb{R})}= \mathrm{SO(3)}\oplus \mathbb{K}
\end{eqnarray}
where
\begin{eqnarray}
&&\mathrm{SL(3,\mathbb{R})}=\mbox{Span}\{ \mathcal{H}_i,E^\alpha,E^{-\alpha}\},\ \ \ i=1,2,\
\ \ \alpha=1,2,3,
 \nonumber\\
&&\mathbb{K}=\mbox{Span}\left \{\mathbb{K}_A \right \}=\mbox{Span}\left \{\mathcal{H}_i,
\frac{1}{\sqrt{2}} \,\left (E^\alpha +E^{-\alpha}\right) \right \}, \nonumber\\
&&\mathrm{SO(3)}=\mbox{Span}\left \{t_\alpha \right \}=\mbox{Span}\left \{  \, \left (E^\alpha
-E^{-\alpha}\right )\right \}.
\end{eqnarray}
In the fundamental $3 \times 3$ representation the generators of $\mathrm{SL(3,\mathbb{R})}$
are given by:
\begin{eqnarray}
 \mathcal{H}_1=\frac{1}{\sqrt{2}} \left (
\begin{array}{ccc}
1&0 &0\\
0&-1 &0\\
0&0 &0
\end{array}\right), \ \ \ \ \ \ \ \
\mathcal{H}_2=\frac{1}{\sqrt{6}} \left (
\begin{array}{ccc}
1&0 &0\\
0&1 &0\\
0&0 &-2
\end{array}\right),\nn\\
 E^1= \left (
\begin{array}{ccc}
0&1 &0\\
0&0 &0\\
0&0 &0
\end{array}\right), \ \ \ \ \ \ \ \
 E^2= \left (
\begin{array}{ccc}
0&0 &0\\
0&0 &1\\
0&0 &0
\end{array}\right), \ \ \ \ \ \ \ \
 E^3= \left (
\begin{array}{ccc}
0&0 &1\\
0&0 &0\\
0&0 &0
\end{array}\right),\nn\\
 E^{-1}= \left (
\begin{array}{ccc}
0&0 &0\\
1&0 &0\\
0&0 &0
\end{array}\right), \ \ \ \ \ \ \ \
 E^{-2}= \left (
\begin{array}{ccc}
0&0 &0\\
0&0 &0\\
0&1 &0
\end{array}\right), \ \ \ \ \ \ \ \
 E^{-3}= \left (
\begin{array}{ccc}
0&0 &0\\
0&0 &0\\
1&0 &0
\end{array}\right).
\end{eqnarray}
In \cite{noiconsasha} we considered the differential system
(\ref{daequa}) generated by taking as generating solution in normal
form the following:
\begin{eqnarray}
  \mathbf{Y}_{nf} &=& \left( \nu_1 ,\nu_2,0,0,0 \right),\nonumber\\
   \nu_1 &=& \frac{\omega -\kappa}{4\sqrt{2}} \quad ; \quad \nu_2 =\frac{3\omega +\kappa}{4 \sqrt{6}}
\label{omegakappainnu}
\end{eqnarray}
where $\omega$ and $\kappa$ are just two alternative constant
parameters replacing the components $\nu_{1,2}$ of the normal
solution along the two Cartan generators. Parametrizing the $h \in
\mathrm{SO(3)}$ gauge group element as follows:
\begin{equation}
  h=\exp \left[ \theta _3(t) \, t_3 \right] \, \exp \left[ \theta _2(t) \, t_2 \right]
  \, \exp \left[ \theta _1(t) \, t_1 \right]
\label{compensah}
\end{equation}
we found in \cite{noiconsasha} that the differential system
(\ref{daequa}) takes the form
\begin{eqnarray}
\nonumber 0&=&{\dot{\theta} _3}(t)-\ft 14\omega \,
          \sin (2\,\theta _3(t)), \\
\nonumber 0&=& 8\,{\dot{\theta} _2}(t) - \left[ \kappa  + \omega
\,\cos (2\,\theta _3(t)) \right] \,
    \sin (2\,\theta _2(t)), \\ \nonumber
0&=& 16{\dot{\theta} _1}(t) + [ \kappa  + \kappa \,\cos (2\,{{\theta
}_2}(t)) + \omega [\cos(2\theta_2(t)) - 3]\cos(2\theta_3(t))]
\,\sin (2\,{{\theta }_1}(t)) - \\ && -8\,\omega \,{\sin^2
({{\theta }_1}(t))}\,\sin ({{\theta }_2}(t))\,
    \sin (2\,{{\theta }_3}(t)).
    \label{A2thetas}
\end{eqnarray}
This system has the requested triangular property and the first two
equations are immediately solved by
\begin{eqnarray}
\theta_3(t) & = & -\arccos \left[- \frac{e^{\frac{\omega \,{{\lambda }_3}}{2}}}
      {{\sqrt{e^{t\,\omega } + e^{\omega \,{{\lambda }_3}}}}}  \right],\nonumber\\
\theta_2(t) & = &  -\arctan {\sqrt{\frac{e^
         {\frac{t\,\left( \kappa  + \omega  \right)  + {{\lambda }_2}}{2}}}{e^
          {t\,\omega } + e^{\omega \,{{\lambda }_3}}}}}
\label{th23}
\end{eqnarray}
where $\lambda_{2,3}$ are two integration constants.
The third equation of (\ref{A2thetas}) was also solved in \cite{noiconsasha} with some
ad hoc elaborations and in this way we obtained a general integral depending
on the five integration constants
$\{\omega,\kappa,\lambda_1,\lambda_2,\lambda_3\}$. This is obviously
consistent with the full integrability of the system demonstrated by
the Lax pair representation, but still required some ad hoc
elaborations. Furthermore for higher rank  algebras like for instance
$A_3$, various experiments showed that although the system
(\ref{daequa}) can be arranged to triangular for a few rotations, it
is not possible to make it completely triangular if we include
rotations along all the generators of $\mathrm{H}$. These limitations
are now completely overcome by the general integration formulae which
we present in the next section.
\section{The general integration algorithm.}
For maximally split cosets $\mathrm{U/H}$, choosing  a solvable coset
representative $\mathbb{L}(\phi)$, there is an established algorithm
\cite{kodama} to construct the general solution of eq. \p{LaxEq}.
For the maximally non compact real section of any simple Lie algebra $\mathbb{U}$
there always exists a permutation matrix $\mathcal{O}$ such that
in the Weyl basis the Lax pair representation \p{LaxEq} can be brought to
the following general form
\begin{eqnarray}
\label{Lax}
\frac{d}{dt} L=\left [L \, , \, P\right ]
\label{newLax}
\end{eqnarray}
where
\begin{eqnarray}
L=\mathcal{O}V\mathcal{O}^T,\ \ \ \ \
P=\mathcal{O}W\mathcal{O}^T
\label{triangabasa}
\end{eqnarray}
and the matrix $P$ is given by the following  projection of $L$
\begin{eqnarray}
 P=\Pi (L):=L_{>0}-L_{<0},
 \label{Lprojection}
\end{eqnarray}
$L_{>0~(<0)}$ denoting the strictly upper (lower) triangular
part of the $n \times n$ matrix $L$. The existence of the matrix $\mathcal{O}$, such that
eq. (\ref{Lprojection}) holds true is guaranteed by a theorem proved
in Helgason's book \cite{Helgason}, by the maximally split nature of
the coset $U/H$ and by the condition of solvability
(\ref{solvocondo}) imposed on the coset representative. The quoted
theorem states that for \textit{any linear representation of a solvable
algebra} there always exist a basis where all generators are upper
triangular matrices. Invoking this theorem for the solvable algebra $Solv(\mathrm{U/H})$ generated
by $\left \{\mathcal{H}_i \, , \, E^\alpha \right \} $ we conclude
that we can always bring the semisimple operators $\mathcal{H}_i$
to diagonal form and the nilpotent operators $E^\alpha$ to strictly
upper triangular form. In the same basis the step-down operators
$E^{-\alpha}$ which are the adjoints of the corresponding
$E^{\alpha}$ are necessarily strictly lower triangular. Since the
connection $W$ is by definition given by $W=t_\alpha \, W^\alpha
= W^\alpha \left( E^\alpha - E^{-\alpha}\right)  $, while $V$ is
given by $V=V^i \, \mathcal{H}_i + V^\alpha \, \frac{1}{\sqrt{2}} \left( E^\alpha - E^{-\alpha}\right)
$, we see that in the basis where $E^\alpha$ is upper triangular and
$\mathcal{H}_i$ diagonal, eq. (\ref{Lprojection}) is nothing else but
the solvability condition (\ref{solvocondo}).
\par
This established, we can proceed to apply the integration algorithm \cite{kodama} in order to write
the general integral of the Lax equation \p{Lax}. Actually this is nothing else
than an instance of the inverse scattering method. Indeed
equation \p{Lax} represents the compatibility condition for the
following linear system exhibiting the iso-spectral property of $L$
\begin{eqnarray}
\label{LaxIs}
L\Psi=\Psi \Lambda,\nn\\
\frac{d}{dt} \Psi=-P\Psi
\end{eqnarray}
where $\Psi(t)$ is the eigenmatrix, namely the matrix whose $i$-th row is
the eigenvector $\varphi(t,\lambda_i)$ corresponding to the eigenvalue $\lambda_i$ of the Lax operator
$L(t)$ at time $t$ and $\Lambda$ is the diagonal
matrix of eigenvalues, which are constant throughout the whole time flow:
\begin{eqnarray}
\Psi&=&\left [\varphi(\lambda_1),\dots,\varphi(\lambda_n)]\equiv[\varphi_i(\lambda_j)\right ]_{1\leq
i,j\leq n},\nonumber\\
\Psi^{-1}&=&\left [\psi(\lambda_1),\dots,\psi(\lambda_n)]^T\equiv[\psi_j(\lambda_i)\right ]_{1\leq
i,j\leq n},\nonumber\\
 \Lambda&=& \mathrm{diag}\left (\lambda_1,\dots, \lambda_n \right ).
 \end{eqnarray}
 The solution of \p{LaxIs} for the Lax operator  is given by  the following explicit  form
 of the matrix elements:
\begin{eqnarray}
\label{sol}\left[ L(t) \right]_{ij}=\sum_{k=1}^n \lambda_k
\varphi_i(\lambda_k,t) \psi_j(\lambda_k,t)~ .
\end{eqnarray}
The eigenvectors of the Lax operator at each instant of time, which define the eigenmatrix $\Psi(t)$, and the
columns of the its inverse $\Psi^{-1}(t)$, are expressed in closed form in terms of
the initial data  at some conventional instant of time, say at
$t=0$.
\par
Explicitly we have:
\begin{eqnarray}
\varphi_i(\lambda_j,t)&=&\frac{e^{-\lambda_j
t}}{\sqrt{D_i(t)D_{i-1}(t)}} \, \mathrm{Det} \, \left ( \begin{array}{cccc}
c_{11}&\dots &c_{1,i-1}& \varphi_1^0(\lambda_j)\\
\vdots&\ddots&\vdots&\vdots\\
c_{i1}&\dots &c_{i,i-1}& \varphi_i^0(\lambda_j)\\
\end{array}\right
),\nn\\
\psi_j(\lambda_i,t)&=&\frac{e^{-\lambda_i t}}{\sqrt{D_j(t)D_{j-1}(t)}}
\, \mathrm{Det} \, \left ( \begin{array}{ccc}
c_{11}&\dots &c_{1,j}\\
\vdots&\ddots&\vdots\\
c_{j-1,1}&\dots &c_{j-1,j}\\
 \psi_1^0(\lambda_i)&\dots &\psi_j^0(\lambda_i)
\end{array}\right
) \end{eqnarray}
where the time dependent matrix $c_{ij}(t)$ is defined below:
\begin{eqnarray}
c_{ij}(t)=\sum_{k=1}^N e^{-2\lambda_k t}
\varphi_i^0(\lambda_k)\psi_j^0(\lambda_k),
\end{eqnarray}
and
\begin{eqnarray}
  \varphi_i^0(\lambda_k) & := & \varphi_i(\lambda_k,0) \nonumber\\
  \psi_i^0(\lambda_k) & := &\psi_i(\lambda_k,0)
\label{eigenvecat0}
\end{eqnarray}
are the eigenvectors and their adjoints calculated at $t=0$.
 These constant vectors constitute the initial data of the
problem and provide the integration constants. Finally $D_k(t)$ denotes the determinant of the $k\times k$ matrix
with entries $c_{ij}(t)$
 \begin{eqnarray}
 D_k(t)=\mathrm{Det} \Biggr [ \Bigr ( c_{ij}(t) \Bigr )_{1\leq i,j \leq k} \Biggr
 ].
 \label{Ddefi}
\end{eqnarray}
Note that $c_{ij}(0)=\delta_{ij}$ and $D_k(0)=1$ as well as $D_0(t):=1$.
\subsection{Parametrization of the initial data}
Let us now reconsider the integration formulae presented in the
previous subsection keeping in mind that, at any instant of time $t$,
the Lax operator $L(t)$ is an element of the Lie algebra $\mathbb{U}$
lying in the orthogonal complement $\mathbb{K}$ to the subalgebra
$\mathbb{H}$. This Lie algebra element is expressed in the chosen
$n$--dimensional representation. Diagonalizing $L(t)$ means
nothing else than bringing it inside the Cartan subalgebra $\mathrm{CSA} \subset \mathbb{K} \subset
\mathbb{U}$, which can always be performed by conjugation with an
element of the maximal compact subgroup. Hence at every instant of time
there exists an element $h(t) \in \exp \left [ \mathbb{H} \right ]$
such that:
\begin{equation}
\exists  \,  h(t) \in \exp \left [ \mathbb{H} \right ]  \, \backslash
\, h^{-1}(t) \, L(t) \, h(t) = \mathcal{C}_0 \, \in \, \mathrm{CSA}.
\label{CSAelement}
\end{equation}
The $n \times n$ matrix   $\mathcal{C}_0$ is by construction diagonal
and has therefore on its diagonal the above mentioned constant eigenvalues
$\lambda_i$. What are these? They are just the $n$--weights of the
chosen $n$--dimensional representation evaluated on the Cartan
element $\mathcal{C}_0$. Calling $\mathbf{w}_i$ the weights, we just
have:
\begin{eqnarray}
  \lambda_i &=& \mathbf{w}_i\left( \mathcal{C}_0 \right) =
  \mathbf{w}_i\cdot \overrightarrow{\mathbf{\nu}} \quad (i=1,\dots , n)
\label{wunu}
\end{eqnarray}
having denoted by the $r$-vector $\overrightarrow{\mathbf{\nu}}$ the
$r$ components of the Cartan element $\mathcal{C}_0$ in a basis of
the Cartan subalgebra:
\begin{equation}
  \mathcal{C}_0 = \nu_i \, \mathcal{H}^i      \quad i = 1, \dots , r
  = \mathrm{rank} (\mathbb{U}).
\label{C0}
\end{equation}
Equation (\ref{CSAelement}) is in particular true at the initial time
$t=0$, where there exists some constant $h_0 \in  \exp \left [\mathbb{H} \right ]$ such that
\begin{equation}
  h_0^{-1} \, L(0) \, h_0 =
  \mathcal{C}_0 =\mathrm{diag}\left( \lambda_1 \, \dots, \, \lambda_n\right).
\label{agnisco}
\end{equation}
At this point all items entering the solution algorithm described in the previous section obtain
their proper group-theoretical interpretation. The diagonal
eigenvalue matrix $\Lambda$ is nothing else but the $\mathrm{CSA}$ element
$\mathcal{C}_0$ evaluated in the chosen $n$--dimensional
representation. The eigenmatrix $\Psi(0)$ which provides the initial
data of the problem is nothing else but the $H$--group element $h_0$.
This explains what we anticipated in section \ref{anticipazia} when we stated
that the initial data are a pair of CSA  algebra element and a H-subgroup element.
We still have to verify the other properties relative to the Weyl
group. These are also easily understood. By definition the Weyl group
$\mathcal{W}$ is that discrete subgroup of $\mathrm{H}$ which maps
the Cartan subalgebra into itself:
\begin{equation}
  \forall w\in\mathcal{W}, \, \, \forall \mathcal{C} \in \mathrm{CSA}
  \quad w^{-1}\cdot \mathcal{C} \cdot w \, \in \, \mathrm{CSA}.
\label{orpo}
\end{equation}
Hence if we choose $h_0$ inside the Weyl group the initial value of
the Lax operator $L_0 =L(0)$ remains a diagonal matrix in the Cartan
subalgebra. If $L_0$ is diagonal the connection $P(0) =0$ vanishes at
the origin and from Lax equation we obtain
\begin{equation}
  \frac{d}{dt} \, L |_{t=0} = 0
\label{implica}
\end{equation}
which implies that the solution is constant throughout the whole
flow: $L(t) = L(0)$. This suffices to prove eq. (\ref{Lweylproperty}).
On the other the so called sorting property, namely the statement (\ref{pminftylimits})
was proved by Kodama et al in \cite{kodama}. As already anticipated in
section \ref{anticipazia} eq. (\ref{Lweylproperty}), together with the
iso-spectral property, namely with the form of the eigenvalues
(\ref{wunu}) implies that eq. (\ref{topcharge}) has also to be true.
Finally the independence of the charge $\Omega$ from the eigenvalues,
namely from the choice of the $\mathrm{CSA}$ element $\mathcal{C}_0$,
together with eq. (\ref{Lweylproperty}) implies that the map induced
by the solution algorithm is indeed of the type described in
eq. (\ref{Pmappa}).
\section{The $A_3$ example with numerical plots  }
In order to exemplify the value and the use of the general
integration algorithm, we will apply it to the case of the manifold
\begin{equation}
  \mathcal{M}_3 \equiv \frac{\mathrm{SL(4,\mathbb{R})}}{\mathrm{O(4)}}
\label{M3manif}
\end{equation}
which is maximally split of rank $r=3$ and corresponds to the Lie
algebra $A_3$. As in the previous $A_2$ case we can use the fundamental
representation as working representation for the Lax operator, which
means that we deal with $4 \times 4 $ matrices. The dimension of the
coset is $9$ and there are $6$ positive roots:
\begin{equation}
\begin{array}{ccccccc}
 \alpha_1 & = &\left \{ {\sqrt{2}},0,0 \right\} & ; &
\alpha_2 & = &\left  \{ -\frac{1}{{\sqrt{2}}}  ,
  -{\sqrt{\frac{3}{2}}},0 \right \},\\
  \alpha_3 & = &\left \{
  0,{\sqrt{\frac{2}{3}}},\frac{2}{{\sqrt{3}}}\right \} & ; &
  \alpha_4 & = & \left \{
  \frac{1}{{\sqrt{2}}},-{\sqrt{\frac{3}{2}}},0 \right \},\\
  \alpha_5 & = &\left  \{ - \frac{1}{{\sqrt{2}}}  ,
  - \frac{1}{{\sqrt{6}}}  ,
  \frac{2}{{\sqrt{3}}}\right \} & ; &
  \alpha_6 & = &\left  \{ \frac{1}{{\sqrt{2}}},
  - \frac{1}{{\sqrt{6}}}  ,
  \frac{2}{{\sqrt{3}}}\right \}. \
\end{array}
\label{ruttine}
\end{equation}
The matrices representing the Cartan generators well adapted to the
representation (\ref{ruttine}) are the following ones:
\begin{equation}
  \begin{array}{cccccccc}
    \mathcal{H}_1 & = & \left(\matrix{ \frac{1}{{\sqrt{2}}} & 0 & 0 & 0 \cr 0 & -
     \frac{1}{{\sqrt{2}}}  & 0 & 0 \cr 0 & 0 &
   0 & 0 \cr 0 & 0 & 0 & 0 \cr  } \right)  & ; & \mathcal{H}_2 & = & \left(\matrix{ - \frac{1}{{\sqrt{6}}}
      & 0 & 0 & 0 \cr 0 & - \frac{1}
     {{\sqrt{6}}}  & 0 & 0 \cr 0 & 0 & {\sqrt
      {\frac{2}{3}}} & 0 \cr 0 & 0 & 0 & 0 \cr  } \right) & ; \\
        \mathcal{H}_3 & = & \left(\matrix{ \frac{1}{2\,{\sqrt{3}}} & 0 & 0 & 0 \cr 0 &
    \frac{1}{2\,{\sqrt{3}}} & 0 & 0 \cr 0 & 0 & \frac{1}
   {2\,{\sqrt{3}}} & 0 \cr 0 & 0 & 0 & \frac{-{\sqrt{3}}}
   {2} \cr  } \right) & ; & \null & \null & \null \
  \end{array}
\label{rotta}
\end{equation}
while the $6$ step-up operators associated  with the positive roots are
given by:
\begin{equation}
  \begin{array}{ccccccc}
    E^{\alpha_1} & = & \left(\matrix{
   0 & 1 & 0 & 0 \cr 0 & 0 & 0 & 0 \cr 0 & 0 & 0 & 0 \cr
   0 & 0 & 0 & 0 \cr  } \right)  & ; & E^{\alpha_2} & = & \left(\matrix{
   0 & 0 & 0 & 0 \cr 0 & 0 & 1 & 0 \cr 0 & 0 & 0 & 0 \cr
   0 & 0 & 0 & 0 \cr  } \right),  \
  \end{array}
\label{ep12}
\end{equation}
\begin{equation}
  \begin{array}{ccccccc}
    E^{\alpha_3} & = & \left( \matrix{
   0 & 0 & 0 & 0 \cr 0 & 0 & 0 & 0 \cr 0 & 0 & 0 & 1 \cr
   0 & 0 & 0 & 0 \cr  }\right)  & ; & E^{\alpha_4} & = & \left(\matrix{
   0 & 0 & 1 & 0 \cr 0 & 0 & 0 & 0 \cr 0 & 0 & 0 & 0 \cr
   0 & 0 & 0 & 0 \cr  } \right),  \
  \end{array}
\label{ep34}
\end{equation}
\begin{equation}
  \begin{array}{ccccccc}
    E^{\alpha_5} & = & \left(\matrix{
   0 & 0 & 0 & 0 \cr 0 & 0 & 0 & 1 \cr 0 & 0 & 0 & 0 \cr
   0 & 0 & 0 & 0 \cr  } \right)  & ; & E^{\alpha_6} & = & \left( \matrix{
   0 & 0 & 0 & 1 \cr 0 & 0 & 0 & 0 \cr 0 & 0 & 0 & 0 \cr
   0 & 0 & 0 & 0 \cr }\right).  \
  \end{array}
\label{ep56}
\end{equation}
The step-down operators associated with the negative roots are just
the transposed of the corresponding step-up operators.
\par
The $\mathrm{H}$ subgroup is $\mathrm{O(4)}$, namely the set of all
orthogonal $4 \times 4$--matrices. The Weyl group $ \mathcal{W}
\subset \mathrm{O(4)} \subset \mathrm{SL(4,\mathbb{R})}$ is
$S_4$ and its elements are represented by the orthogonal matrices
associated with the $4!$ permutations $P \in S_4$ in the following
way:
\begin{eqnarray}
  \left( O_P\right) _{i,P(i)}&=&1, \nonumber\\
  \left( O_P\right) _{ij} & = & 0 \quad \mbox{otherwise}.
\label{Opmatri}
\end{eqnarray}
So for instance the cyclic permutation
\begin{equation}
  P \quad = \quad \begin{array}{cccc}
    1 & 2 & 3 & 4 \\
    \downarrow & \downarrow & \downarrow & \downarrow \\
    4 & 1 & 2 & 3 \
  \end{array}
\label{uppa}
\end{equation}
is represented by the matrix
\begin{equation}
O_P =  \left(\begin{array}{cccc}
  0 & 0 & 0 & 1 \\
  1 & 0 & 0 & 0 \\
  0 & 1 & 0 & 0 \\
  0 & 0 & 1 & 0
\end{array} \right).
\label{cyclic}
\end{equation}
Having defined the model we can now study the solutions produced by
the general integration algorithm implementing it on a computer and
choosing numerical values for the initial data.
\par
Let us begin with the choice of the Cartan subalgebra element
$\mathcal{C}_0$.
Setting:
\begin{equation}
   {{{{\nu }_1}} =
    {-\left( \frac{{{\lambda }_1} + {{\lambda }_2} +
          2\,{{\lambda }_3}}{{\sqrt{2}}} \right) }} \quad ; \quad
  {{{{\nu }_2}}=
    {\frac{{{\lambda }_1} + 3\,{{\lambda }_2}}
      {{\sqrt{6}}}}} \quad ; \quad {{{{\nu }_3}}=
    {\frac{-2\,{{\lambda }_1}}{{\sqrt{3}}}}}.
\label{pop}
\end{equation}
We obtain:
\begin{equation}
  \mathcal{C}_0 \equiv \sum_{i=1}^{3} \, \nu_i \, \mathcal{H}_i \, =
  \, \left(\matrix{ -{{\lambda }_1} - {{\lambda }_2} -
   {{\lambda }_3} & 0 & 0 & 0 \cr 0 & {{\lambda }_
    3} & 0 & 0 \cr 0 & 0 & {{\lambda }_
    2} & 0 \cr 0 & 0 & 0 & {{\lambda }_1} \cr  } \right).
\label{C0choice}
\end{equation}
We are free to choose the eigenvalues $\lambda_i$. We exhibit the
solution for the simple choice:
\begin{equation}
  \lambda_i = i \quad \Rightarrow \quad \mathcal{C}_0 =
  \left(\matrix{ -6 & 0 & 0 & 0 \cr 0 & 3 & 0 & 0 \cr 0 & 0 & 2 & 0 \cr
   0 & 0 & 0 & 1 \cr  }  \right).
\label{c0num}
\end{equation}
Next we have to choose the $H$--subgroup element. Our numerical
choice is given by the following orthogonal matrix:
\begin{equation}
  h_0 = \left( \matrix{ \frac{-3 - {\sqrt{3}}}{8} & \frac{-5\,
     \left( -1 + {\sqrt{3}} \right) }{16} & \frac{1 + {\sqrt{3}}}
   {4\,{\sqrt{2}}} & \frac{-1 - 5\,{\sqrt{3}}}{16} \cr \frac{-\left( -1 +
       {\sqrt{3}} \right) }{4\,{\sqrt{2}}} & \frac{{\sqrt{2}} +
     3\,{\sqrt{6}}}{16} & \frac{1 + {\sqrt{3}}}{4} & \frac{-\left( -7 +
       {\sqrt{3}} \right) }{8\,{\sqrt{2}}} \cr \frac{-\left( 3 +
       2\,{\sqrt{3}} \right) }{8\,{\sqrt{2}}} & \frac{7 + 4\,{\sqrt{3}}}
   {16\,{\sqrt{2}}} & \frac{-6 + {\sqrt{3}}}{8} & \frac{-4 + {\sqrt{3}}}
   {16\,{\sqrt{2}}} \cr \frac{-8 + {\sqrt{3}}}{8\,{\sqrt{2}}} & \frac
     {-\left( 10 + {\sqrt{3}} \right) }{16\,{\sqrt{2}}} & -\left( \frac{1}
     {8} \right)  & \frac{11 + 2\,{\sqrt{3}}}{16\,{\sqrt{2}}} \cr  }\right)
\label{okramatra}
\end{equation}
which is just a product of rotations along the six generators of
$\mathrm{O(4)}$ with angles that are multiples of $\pi/3$ or $\pi/4$ in order
to obtain simple but non trivial entries all over the place.
With such initial data the computer programme calculates the Lax
operator $L(t)$ at all times and finds that the limits at $\pm\infty$
are as follows:
\begin{eqnarray}
\lim_{t\rightarrow -\infty} \, L(t) & = & \left( \matrix{
   3 & 0 & 0 & 0 \cr 0 & 2 & 0 & 0 \cr 0 & 0 & 1 & 0 \cr 0 & 0 & 0 &
    -6 \cr  }\right),  \nonumber\\
\lim_{t\rightarrow \infty} \, L(t)  & = & \left( \matrix{ -
    6 & 0 & 0 & 0 \cr 0 & 1 & 0 & 0 \cr 0 & 0 & 2 & 0 \cr 0 & 0 & 0 &
   3 \cr  }\right).
\label{copperfi}
\end{eqnarray}
This means that the image of $h_0 \in \mathrm{O(4)}$ under the map
$\mathcal{P}_K $ of eq. (\ref{Pmappa}) is the permutation displayed below:
\begin{equation}
\mathcal{P}_K  \quad : \quad \frac{\mathrm{O(4)}}{S_4} \, \ni \, h_0 \, \rightarrow \,
\pi_1 \left( h_0 \right)
\, = \,\left(  \matrix{
   0 & 0 & 0 & 1 \cr 0 & 0 & 1 & 0 \cr 0 & 1 & 0 & 0 \cr 1 & 0 & 0 &
   0 \cr  } \right ) \, \in \, S_4.
\label{mappolona}
\end{equation}
The permutation of the eigenvalues occurs through a smooth process
that realizes the billiard bounces and that can be appreciated
through the inspection of plots of the Cartan fields. The most
instructive plots are those of the differential and integrated Cartan
fields. Given the explicit solution for the Lax operator $L(t)$
obtained through implementation of the integration algorithm, let us
define:
\begin{equation}
  \overrightarrow{\chi}_x(t) \, \equiv \, \mbox{Tr} \left[ \mathcal{H}_x \, L(t) \right]
\label{tancarta}
\end{equation}
the components of the tangent vector along the geodesic.
\iffigs
\begin{figure}
\begin{center}
\epsfxsize =6cm
{\epsffile{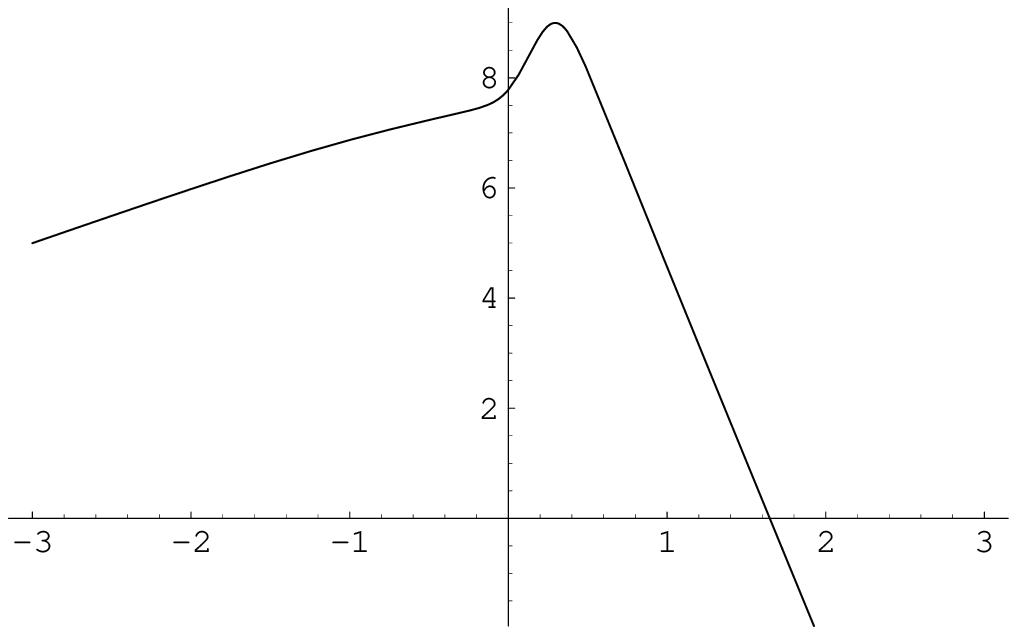}}
\epsfxsize =6cm
{\epsffile{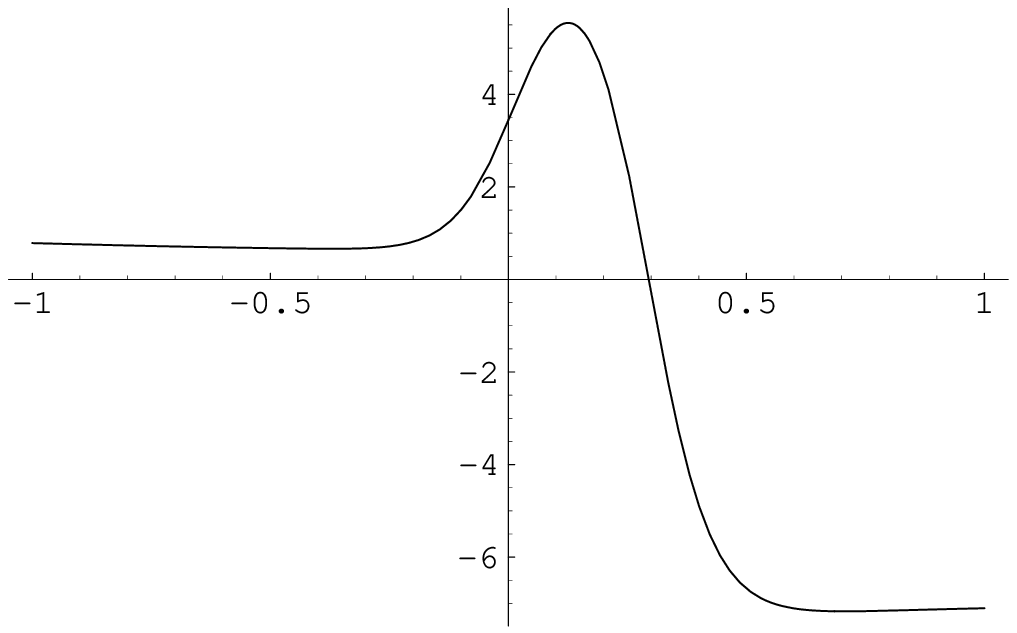}}
%\vskip -8.5cm
\caption{Plots of the differential and integrated Cartan fields $\chi_{\alpha_1}(t)$ and $h_{\alpha_1}(t)$, along
the first simple root of the $A_3$ Lie algebra and for the solution
generated by the initial data $C_0$ and $h_0$, given in eqs.
(\ref{c0num}) and (\ref{okramatra}), respectively. The plot on the left is $h_{\alpha_1}(t)$, while the
plot on the right is $\chi_{\alpha_1}(t)$. We see the bounce at $t= 0.29488$.\label{plottucci1}  }
\hskip 2cm \unitlength=1.1mm
\hskip 1.5cm \unitlength=1.1mm
\end{center}
\end{figure}
\fi
\iffigs
\begin{figure}
\begin{center}
\epsfxsize =6cm
{\epsffile{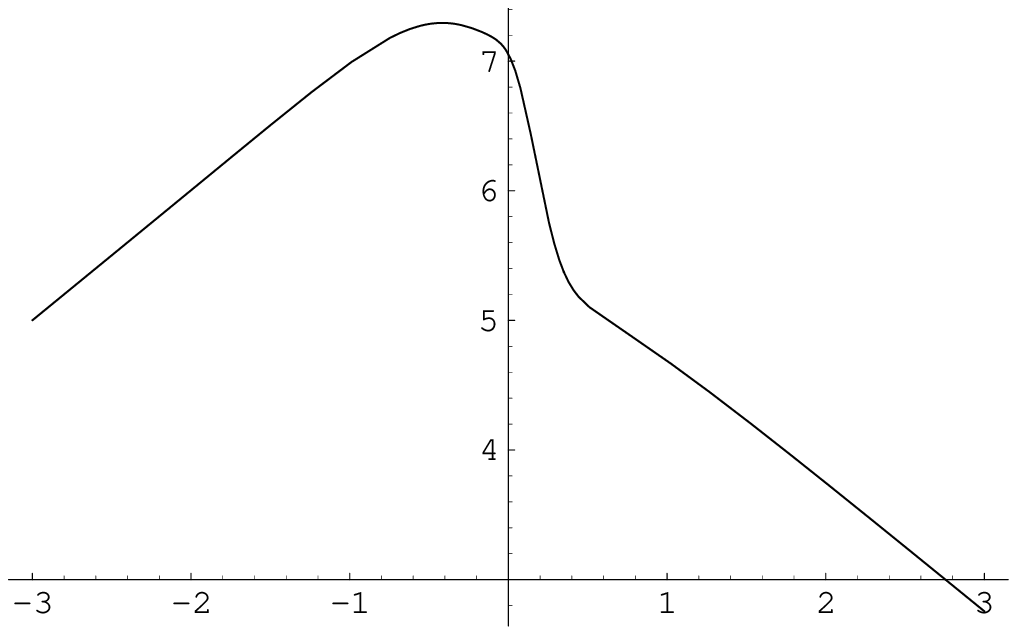}}
\epsfxsize =6cm
{\epsffile{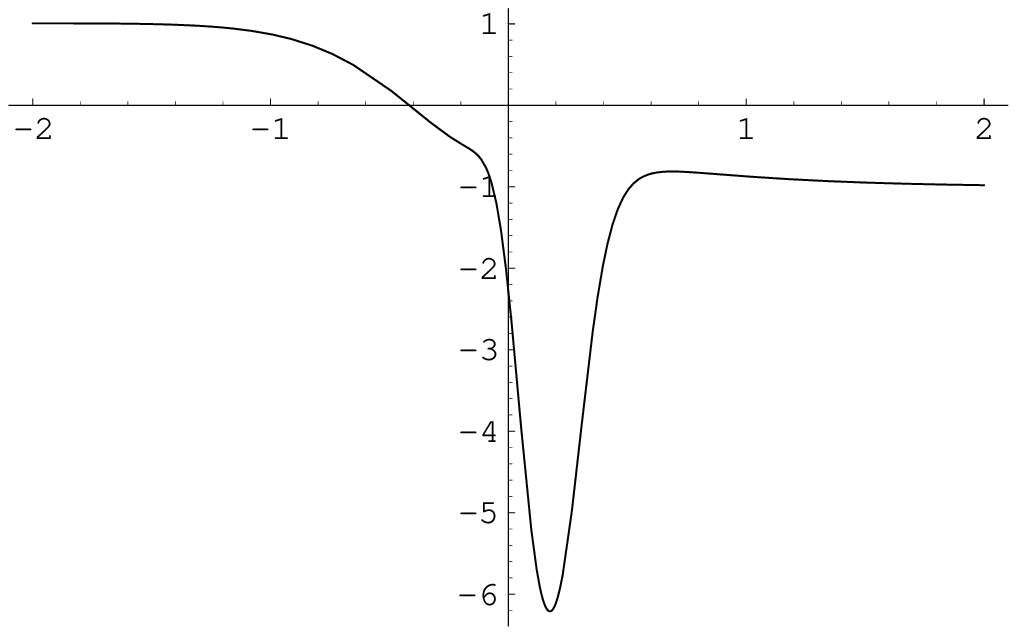}}
%\vskip -8.5cm
\caption{Plots of the differential and integrated Cartan fields $\chi_{\alpha_2}(t)$ and $h_{\alpha_2}(t)$, along
the second simple root of the $A_3$ Lie algebra and for the solution
generated by the initial data $C_0$ and $h_0$, given in eqs.
(\ref{c0num}) and (\ref{okramatra}), respectively. The plot on the left is $h_{\alpha_2}(t)$, while the
plot on the right is $\chi_{\alpha_2}(t)$. We see the bounce at $t= - 0.416$\label{plottucci2}  }
\hskip 2cm \unitlength=1.1mm
\hskip 1.5cm \unitlength=1.1mm
\end{center}
\end{figure}
\fi
\iffigs
\begin{figure}
\begin{center}
\epsfxsize =6cm
{\epsffile{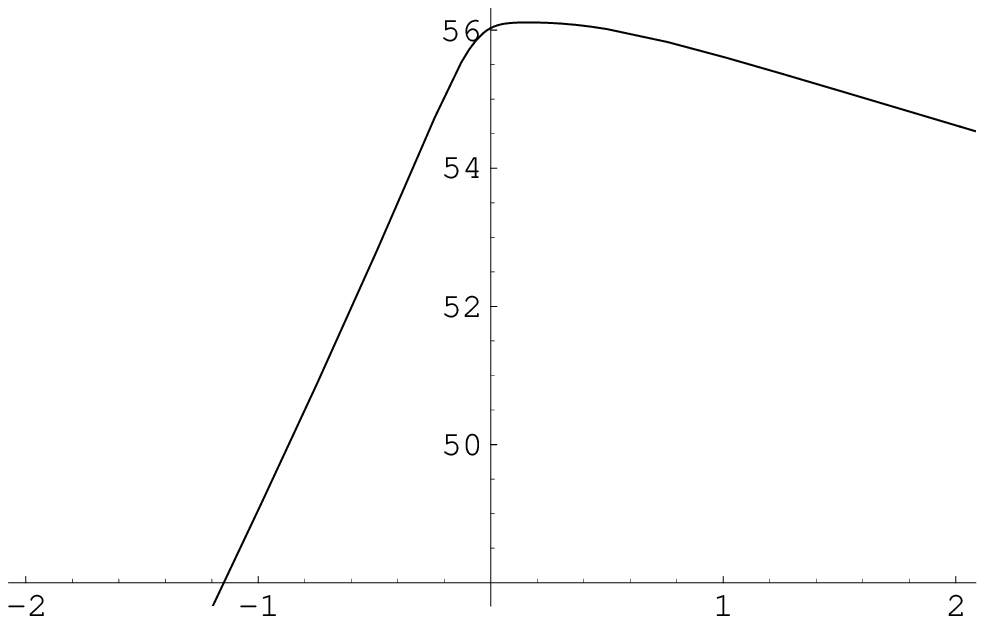}}
\epsfxsize =6cm
{\epsffile{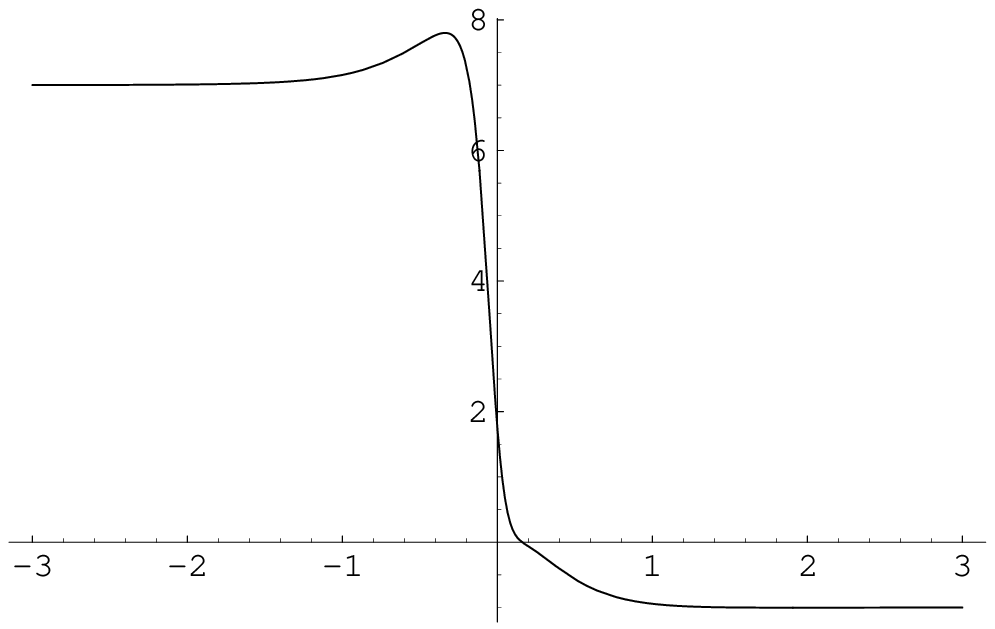}}
%\vskip -8.5cm
\caption{Plots of the differential and integrated Cartan fields $\chi_{\alpha_3}(t)$ and $h_{\alpha_3}(t)$, along
the third simple root of the $A_3$ Lie algebra and for the solution
generated by the initial data $C_0$ and $h_0$, given in eqs.
(\ref{c0num}) and (\ref{okramatra}), respectively. The plot on the left is $h_{\alpha_3}(t)$, while the
plot on the right is $\chi_{\alpha_3}(t)$. We see the bounce at $t= 0.160124$. \label{plottucci3}  }
\hskip 2cm \unitlength=1.1mm
\hskip 1.5cm \unitlength=1.1mm
\end{center}
\end{figure}
\fi
Let us also define the projection of this tangent vector along the simple roots
\begin{equation}
  \chi_{\alpha_i}(t) \, = \, \overrightarrow{\chi}(t) \cdot \alpha_i
  \quad ; \quad i=1,2,3 ~.
\label{chialongsimple}
\end{equation}
Finally the Cartan fields, that in supergravity correspond to the
cosmological scale factors and to the dilatons \footnote{See
\cite{noiconsasha,Weylnashpaper} for detailed explanations about the oxidation mechanism  that lifts the
$\sigma$--model solutions to full--fledged solutions of supergravity theory in higher dimensions.
See also \cite{noiKacmodpaper}
and \cite{noipaintgroup} for the relation between the $\sigma$--model and $D=4$ supergravity with all
number of supercharges.}, are defined by simple integration:
\begin{equation}
  h_{\alpha_i}(t) \, \equiv \, \int_{t_0}^{t} {\chi}_{\alpha_i} \left( \ell\right)
  \, d\ell
  \quad ; \quad i=1,2,3 ~.
\label{integratto}
\end{equation}
In fig.s \ref{plottucci1},\ref{plottucci2},\ref{plottucci3} we
present, for the considered explicit solution the plots of
$\chi_{\alpha_i}(t)$ and $h_{\alpha_i}(t)$.
Billiard bounces correspond to maxima or minima of the simple root
integrated Cartan fields $h_{\alpha_i}(t)$ or, equivalently to zeros  of the
differential ones $\chi_{\alpha_i}(t)$. Inspection of the plots
reveals that the considered solution, characterized by the
topological charge of equation (\ref{mappolona}) admits three bounces,
respectively located at:
\begin{eqnarray}
&& t= 0.160124 \quad; \quad
t= - 0.416 \quad ; \quad
t= 0.29488 ~.
\label{bouncioni}
\end{eqnarray}
\section{Billiard interpretation}
Having in mind the explicit example we have just illustrated we can
now emphasize the physical \textit{billiard interpretation} of the
mathematical results we obtained.
\par
 At those times when the Lax operator $L(t)$ lies in the Cartan subalgebra,
 the fictitious cosmic ball moves on a straight line, the components
 of $L(t)$ along the CSA generators $\mathcal{H}_x$ being the
 components of its velocity:
\begin{equation}
  \overrightarrow{v}_x (t) = \mbox{Tr} \left [ \mathcal{H}_x L(t) \right ].
\label{velocity}
\end{equation}
This means that the multidimensional Universe has a constant rate
(positive or negative) exponential expansion (contraction) in the
corresponding dimensions. The asymptotic theorem states that for
finite Lie algebras, namely as long as we confine our attention to
time-dependent backgrounds that can be  retrieved by oxidation from $D=3$ supergravity
(see \cite{noiconsasha,Weylnashpaper,noipaintgroup,noiKacmodpaper}),
the Universe is in such a \textit{constant velocity state} at
asymptotically early and asymptotically late times. Furthermore given
the initial state:
\begin{equation}
  |\overrightarrow{v}, - \infty >
\label{iniziale}
\end{equation}
at $t=-\infty$ (Big Bang time) the entire cosmic
evolution can be seen as a smooth process that brings the Universe
to another such state:
\begin{equation}
 |\overrightarrow{v}^\prime, + \infty >.
\label{finale}
\end{equation}
The surprising result is that there is a finite countable numbers of
such possible states, as many as the elements of the Weyl group $
\mathcal{W}$. Indeed the final Universe velocity is necessarily the
image of its initial one under the action of some element of $
\mathcal{W}$. The transition from one state to the other occurs
through one or several bounces on the dynamical walls of the Weyl
chamber that as we have already emphasized in previous papers raise
and decay smoothly.
%%%%%%%%%%%%%%%%%%%%%%%%%%%%%%%%%%%
\section{Conclusions and Perspectives}
We could now show explicit analytic formulae for the general integral of
simple models like the $A_2$ model. It is very easy to obtain them by
means of computer codes written in MATHEMATICA but they are very lengthy and
clumsy to be displayed so that we prefer not to. The relevant point
is that the integration algorithm is fully explicit and we hope we
have been able to illustrate this fact. This enables us to apply our
method to any case of interest, as long as the target manifold is
maximally split. We can also generate, by means of the \textit{paint group
rotations} entire classes of solutions in the non maximally split
case. However they do not yet constitute a general integral in these
cases. The open problem is therefore that of extending the results
presented in this article in the following  directions:
\begin{enumerate}
  \item For the non maximally split non compact cosets appearing in
  supergravity with non maximal SUSY.
  \item For the infinite dimensional Ka\v c--Moody algebras generated
  by reduction to $D=2$ and $D=1$. This is particularly important in
  view of the results we have achieved on asymptotic regimes. It
  seems quite clear that the chaotic Kasner behaviour predicted
  around initial or final singularities must be related to the
  essential change in structure of the Weyl group for Ka\v c--Moody
  algebras. The presence of infinitely many roots and moreover the
  presence of space--like or null--like roots makes the Weyl group
  much more complicated and apt to create smooth solutions with
  singular behaviour if there is an extension of our main results to
  the Ka\v c--Moody case.
  \item For $\sigma$--models deformed by the presence of a potential,
  like it happens when fluxes are turned on and one considers gauged
  supergravity rather than ungauged supergravity.
  \item Generalization of the results obtained in this paper  to the case of
one-dimensional super-coset sigma models, which comprise both bosonic
and fermionic fields.
\end{enumerate}
On all of these four issues we are presently working and hope to be
able to present new conclusions soon.

~{}

{\bf Acknowledgements}
We would like to thank V.V. Gribanov for the collaboration at the earlier
stage of this investigation. A.S. thanks Torino University for the kind
hospitality at the final stage of this work. The work of A.S. was partially
supported by the RFBR-DFG Grant No. 04-02-04002, DFG Grant 436 RUS 113/669-2,
NATO Grant PST.GLG.980302 and by the Heisenberg-Landau program.

\end{document}